\documentclass[aps,prb,twocolumn,floatfix,superscriptaddress]{revtex4-2}
\usepackage{graphicx,amsmath,amsfonts,mathrsfs,bbold,dsfont}

\begin{document}

\title{The origin of ferromagnetic interactions in NaMnCl$_3$: how the response theory reconciles with Goodenough-Kanamori-Anderson rules}

\author{I. V. Solovyev}
\email{SOLOVYEV.Igor@nims.go.jp}
\affiliation{National Institute for Materials Science, MANA, 1-1 Namiki, Tsukuba, Ibaraki 305-0044, Japan}
\affiliation{Institute of Metal Physics, S. Kovalevskaya str. 18, 620108 Ekaterinburg, Russia}
\author{A. V. Ushakov}
\affiliation{Institute of Metal Physics, S. Kovalevskaya str. 18, 620108 Ekaterinburg, Russia}
\author{S. V. Streltsov}
\affiliation{Institute of Metal Physics, S. Kovalevskaya str. 18, 620108 Ekaterinburg, Russia}

\date{\today}

\date{\today}
\begin{abstract}
The on-site Coulomb repulsion $U$ is the key ingredient for describing the magnetic properties of Mott insulators, leading to a popular believe that many limitations of the density-functional theory based methods can be cured by artificially incorporating such on-site interactions for localized electrons in the model form. The layered antiferromagnet NaMnCl$_3$ reveals quite a different story: while the Coulomb $U$ on the Mn sites controls the strength of antiferromagnetic superexchange interactions, an equally important parameter is the Stoner coupling ${\cal I}_{\rm Cl}$ on the ligand sites. The latter is responsible for large ferromagnetic contributions to the interatomic exchange interactions, which in NaMnCl$_3$ nearly cancel the effect of the superexchange interactions. Although such behavior is anticipated from the phenomenological Goodenough-Kamanori-Anderson rules, the quantitative description of the ligand-related contributions remains disputable. Considering NaMnCl$_3$ as an example, we discuss how they can be generally taken into account in the linear response theory to regain the dependence of the exchange interactions on ${\cal I}_{\rm Cl}$. The problem is complicated by the fact that, for the nearly filled Cl $3p$ shell, the parameters ${\cal I}_{\rm Cl}$ are sensitive to the model assumptions.
\end{abstract}

\maketitle

\par \textit{Introduction}. -- In most cases, the exchange interactions between half-filled $3d$ ions are expected to be antiferromagnetic due to the superexchange mechanism~\cite{Anderson}. Such situation is indeed realized in the canonical MnO~\cite{MnO}, the multiferroic BiFeO$_3$~\cite{BiFeO3} and MnI$_2$~\cite{MnI2}, the lithium-ion batteries cathode LiMnPO$_4$~\cite{LiMnPO4}, the flocculant compound FeCl$_3$~\cite{FeCl3}, and other prominent magnetic materials composed from the Mn$^{2+}$ or Fe$^{3+}$ ions. However, when the ${\rm T}$-${\rm L}$-${\rm T}$ exchange path between two transition-metal (${\rm T}$) sites connected by a ligand (${\rm L}$), becomes close to $90^{\circ}$, the situation may be less certain as there are several mechanisms operating in opposite directions and partially cancelling each-other. Besides antiferromagnetic superexchange, such mechanisms typically involve ferromagnetic Hund's coupling on the ligand states. In fact, by summarising the famous Goodenough-Kamanori-Anderson (GKA) rules for transition-metal compounds, Junjiro Kanamori has concluded that the sign of exchange interaction in this $90^{\circ}$-case is ``uncertain''~\cite{Kanamori_GKA}.

\par
\begin{figure}[b]
\begin{center}
\includegraphics[width=0.48\textwidth]{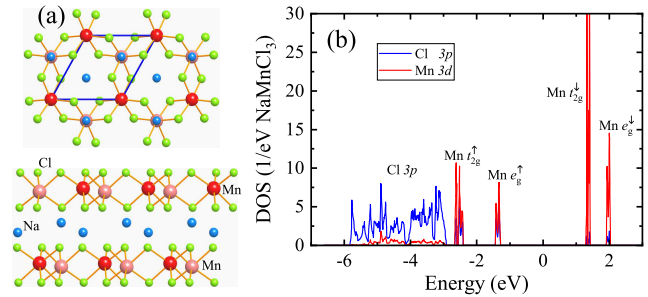}
\end{center}
\caption{(a) Crystal structure of NaMnCl$_3$: top view (top) and side view (bottom). The unit cell in the basal plane is shown by blue solid lines. Two Mn sites in the primitive cell are denoted by different colors. (b) Partial densities of Mn $3d$ and Cl $3p$ states, as obtained in the generalized gradient approximation for the antiferromagnetic state. The zero energy is in the middle of the gap between occupied and empty states. The positions of main bands are indicated by symbols.}
\label{fig.str}
\end{figure}

 NaMnCl$_3$, which was recently reinvestigated by Devlin and Cava~\cite{DevlinCava}, is the good example of such uncertainty. It crystallizes in the trigonal structure with the space group $R\overline{3}$ (Fig.~\ref{fig.str}a), which is similar to that of the van der Waals ferromagnet CrI$_3$~\cite{CrI3_Nature}, except that the [MnCl$_3$]$^{-}$ layers are negatively charged and interconnected by the Na$^{+}$ layers. The Mn$^{2+}$ ions form the distorted honeycomb lattice and the Mn-Cl-Mn angle is $96.7^{\circ}$~\cite{NaMnCl3_str}. NaMnCl$_3$ orders antiferromagneticaly below $T_{\rm N} \approx 6.3$ K~\cite{DevlinCava,NaXCl3}. Nevertheless, the Curie-Weiss constant is positive, $\theta_{\rm CW} \approx 4.2$ K, suggesting that some of the exchange interactions are ferromagnetic. This finding is further elaborated by powder neutron diffraction measurements indicating at the layered antiferromagnetic (LA) structure of NaMnCl$_3$ in which the ferromagnetic layers are coupled antiferromagnetically~\cite{neutron}. In the light of these delicate magnetic properties, NaMnCl$_3$ appears to be an interesting system for testing abilities of first-principles methods based on the density-functional theory (DFT).
\noindent
\noindent

\par Construction of the spin models inevitably relies on some assumptions. One of them is the use of magnetic force theorem (MFT) for interatomic exchange interactions~\cite{LKAG1987}, which becomes a popular tool in the DFT community~\cite{IS2003,Kvashnin,Korotin,Yoon}. However, apart from the fundamental limitations of MFT~\cite{BrunoPRL2003,Antropov2006,PRB2021}, most of such calculations consider the exchange processes only between magnetic $3d$ states, where ligands play a role of an effective medium, which only helps to transfer electrons from one ${\rm T}$-site to another. The contributions caused by the magnetic polarization of the ligand states, $m^{z}_{\rm L}$, are typically ignored or treated in a phenomenological way, by associating them with direct exchange integrals~\cite{Ku,Mazurenko,Danis} or considering an \textit{ad hoc} energy gain caused by this $m^{z}_{\rm L}$~\cite{Streltsov2008,JPSJ}. In this article, by employing formally exact technique~\cite{PRB2021}, we explicitly show how the ligand-related contributions to the exchange parameters can be naturally evaluated in the framework of the linear response theory. We further demonstrate that, if defined properly, such exchange interactions depend not only on the Coulomb repulsion $U$ on the Mn sites, but also on the Stoner coupling on the ligand sites, ${\cal I}_{\rm Cl}$, responsible for Hund's first rule.

\par \textit{Theory}. -- In order to evaluate the exchange interactions in DFT, we use the exact expression for the total energy change caused by infinitesimal rotations of spins, which is formulated in terms of the inverse response function~\cite{BrunoPRL2003,Antropov2006,PRB2021}. This energy change can be mapped onto the spin model
\noindent
\begin{equation}
{\cal E} = -\frac{1}{2} \sum_{ij} J_{ij} \boldsymbol{e}_{i} \cdot \boldsymbol{e}_{j}
\label{eq:Heis}
\end{equation}
\noindent with $\boldsymbol{e}_{i}$ being the unit vector along the direction of spin at the $i$th ${\rm T}$-site. The corresponding exchange interactions in the reciprocal space include two contributions~\cite{PRB2021}: $J_{\rm TT}(\boldsymbol{q}) = J_{\rm TT}^0(\boldsymbol{q}) + \Delta J_{\rm TT}(\boldsymbol{q})$, where $J_{\rm TT}^0(\boldsymbol{q})$ are the bare interactions, while
\noindent
\begin{equation}
\Delta J_{\rm TT}(\boldsymbol{q}) = - J_{\rm TL}^0(\boldsymbol{q}) \left[ J_{\rm LL}^0(\boldsymbol{q}) \right]^{-1} J_{\rm LT}^0(\boldsymbol{q})
\label{eq:jl}
\end{equation}
\noindent takes into account the contributions of the ligand states. In these notations, the subscripts ${\rm TT}$, ${\rm TL}$, ${\rm LL}$, and ${\rm LT}$ mean that all the quantities are the matrices with the indices ($\mu$ and $\nu$) running over the ${\rm T}$- and ${\rm L}$-sites in the unit cell. Such $\Delta J_{\rm TT}(\boldsymbol{q})$ is obtained in the adiabatic limit, where the ${\rm L}$-spins instantaneously follow each configuration of the ${\rm T}$-spins~\cite{PRB2021}. The matrix elements of $J_{\mu \nu}^0(\boldsymbol{q})$ ($\mu, \nu \in {\rm T}$ or ${\rm L}$) are expressed in terms of the response tensor ${\mathbb R}^{\uparrow \downarrow}(\boldsymbol{q})$ as
\noindent
\begin{equation}
J_{\mu \nu}^0(\boldsymbol{q}) = \frac{1}{2} m_{\mu}^{z} \left( \left[ {\mathbb R}^{\uparrow \downarrow}(\boldsymbol{q}) \right]^{-1}_{\mu \nu}  + {\cal I}_{\mu} \delta_{\mu \nu} \right) m_{\nu}^{z},
\label{eq:jexactM}
\end{equation}
\noindent where $m_{\mu}$ is the magnetic moment and the Stoner parameter ${\cal I}_{\mu}$ corresponds to the exchange-correlation energy of the form $-\frac{1}{4} {\cal I}_{\mu} \big(m_{\nu}^{z}\big)^2$. Further details can be found in Refs.~\cite{PRB2021,PRB2022}. The real space parameters $J_{ij}$ are obtained by the Fourier transform of $J_{\mu \nu}(\boldsymbol{q})$. The popular alternative to this, formally exact, approach is MFT~\cite{LKAG1987}, which relies on additional approximations and can be justified  only in the long wavelength and strong-coupling limits~\cite{PRB2021}. Furthermore, the contribution $\Delta J_{\rm TT}(\boldsymbol{q})$ is ignored in most of the MFT based calculations or replaced by a semi-empirical term simulating the direct exchange~\cite{Ku,Mazurenko,Danis}. However, $\Delta J_{\rm TT}(\boldsymbol{q})$ can be very important, especially when it is consideerd in the combination with the exact Eq.~(\ref{eq:jexactM}) for the exchange interactions~\cite{PRB2021}.

\par \textit{Results and Discussions}. --  The electronic structure calculations were performed in the generalized gradient approximation (GGA)~\cite{gga-pbe}, for the experimental structure reported in Ref.~\cite{NaMnCl3_str} using Vienna \textit{ab initio} simulation package (VASP)~\cite{vasp}. After that, several tight-binding (TB) models were constructed in the maximally localized Wannier basis~\cite{wannier90}.

\par The electronic structure of NaMnCl$_3$ is relatively simple, consisting of several groups of isolated bands (Fig.~\ref{fig.str}b), which allows us to construct two types of models. The simplest $d$-model includes ten Mn $3d$ bands located near the Fermi level. In this case, the Wannier basis consists of five $3d$ orbitals (labeled by $m$ and $m'$) at each Mn site. The corresponding hopping parameters $\hat{t}_{ij} = [t_{ij}^{mm'}]$, are associated with the matrix elements of the GGA Hamiltonian in the Wannier basis~\cite{wannier90}. Then, $J_{ij}$ can be evaluated in the superexchange approximation for the half-filled Mn $3d$ shell as: $J_{ij} = - \frac{{\rm Tr}_{L}( \hat{t}_{ij} )^2}{U+4J_{\rm H}}$~\cite{Anderson}, where $U$ and $J_{\rm H}$ are the intraatomic Coulomb repulsion and exchange coupling, respectively, and ${\rm Tr}_{L}$ denotes the trace over the orbital indices. In the mean-field approximation, ${\cal I}_{\rm Mn}$ is related to $U$ and $J_{\rm H}$ as ${\cal I}_{\rm Mn} = \frac{U+4J_{\rm H}}{5}$~\cite{Oles,PRB1998}. We treat ${\cal I}_{\rm Mn}$ as an adjustable parameter and ask which ${\cal I}_{\rm Mn}$ would reproduce the experimental $T_{\rm N} = 6.3$ K~\cite{DevlinCava} for the given hopping parameters $\hat{t}_{ij}$. In order to evaluate $T_{\rm N}$, we use the random phase approximation (RPA), as explained in Ref.~\cite{PRM2019}. This yields ${\cal I}_{\rm Mn} \sim 5.5$ eV, which is unrealistically large and, assuming $J_{\rm H}=0.8$ eV~\cite{AZA}, would correspond to nearly bare value of $U \sim 24$ eV. Furthermore, all interactions $J_{ij}$ are antiferromagnetic, which cannot explain the sign of $\theta_{\rm CW}$.

\par
\begin{figure}[t]
\begin{center}
\includegraphics[width=0.48\textwidth]{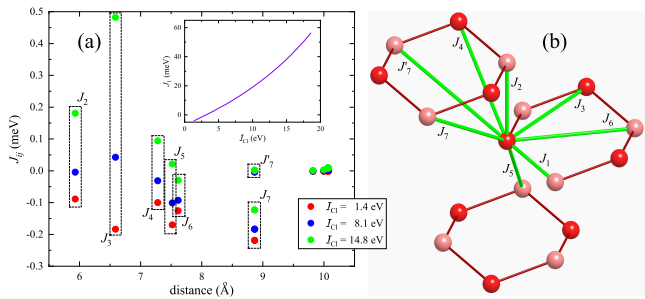}
\end{center}
\caption{(a) Distance dependence of exchange interactions (in GGA) for three values of ${\cal I}_{\rm Cl}$. The inset shows the dependence of $J_1$ on ${\cal I}_{\rm Cl}$. (b) Notations of main exchange interactions.}
\label{fig.J}
\end{figure}

 Thus, it is important to include the Cl $3p$ band explicitly into the model. For these purposes, we construct the TB Hamiltonian in the basis of Mn $3d$ and Cl $3p$ Wannier orbitals, which reproduces the behavior of Mn $3d$ and Cl $3p$ bands in GGA (Fig.~\ref{fig.str}b). In order to evaluate the exchange interactions using Eq.~(\ref{eq:jexactM}), we focus on the antiferromagnetic (A) solution with two Mn spins in the unit cell being antiparallel to each other. The ferromagnetic (F) solution was also investigated and confirmed to provide a similar behavior for the exchange interactions. First, we considered the contributions of the Mn $3d$ states alone, by artificially cutting off all matrix elements of ${\mathbb R}^{\uparrow \downarrow}(\boldsymbol{q})$ associated with the ${\rm L}$ sites, as is typically done in calculations based on MFT. In comparison with the $d$-model, the additional admixture of the Mn $3d$ states into the Cl $3p$ band in the $dp$-model already plays an important role by substantially weakening the antiferromagnetic interactions. For instance, the nearest-neighbor (nn) interaction $J_{1}$ is reduced from $-$$14.4$ till $-$$6.9$ meV (in GGA, see Fig.~\ref{fig.J} for the notations), and similar tendencies are found for other interactions. The complete removal of the Cl $3p$ states (by artificially shifting them downwards by $100$ Ry) yields $J_{1} = -$$4.5$ meV, which can be regarded as the direct Mn-Mn interaction. Therefore, the remaining Mn-Cl-Mn superexchange interactions operating via the Cl $3p$ sites can be estimated in the $d$- and $dp$-model as $-$$9.9$ meV and $-$$2.4$ meV, respectively.
\noindent

\par Up to this point, we obtain a good agreement between the exact theory and MFT~\cite{LKAG1987}. For instance, MFT yields $J_{1} = -$$14.1$ and $-$$6.5$ meV in the $d$- and $dp$-model, respectively, which are indeed very close to the ``exact'' values. This is not surprising, because MFT is justified in the strong coupling limit~\cite{PRB2021}, which is realized for the Mn $3d$ states in NaMnCl$_3$ even at the level of GGA. However, as long as we deal only with the Mn $3d$ states, all interactions remain antiferromagnetic, which is again inconsistent with the type of the magnetic ground state of NaMnCl$_3$ (A instead of LA) as well as the sign of $\theta_{\rm CW}$.

\par The effect of the Cl $3p$ states on the exchange interactions is very dramatic. The Cl $3p$ states are located rather deep in the occupied part and their hybridization with the Mn $3d$ states is not particularly strong. In such a situation, the sub-block $\big[{\mathbb R}^{\uparrow \downarrow}(\boldsymbol{q})\big]_{\rm LL}$ of the response matrix is small (and would completely vanish in the ionic limit, where there is no hybridization with the Mn $3d$ states). Therefore, the inverse matrix $\left[ {\mathbb R}^{\uparrow \downarrow} (\boldsymbol{q}) \right]^{-1}$ will nearly diverge and in the calculations of $J_{ij}^{\phantom{0}} = J_{ij}^0 + \Delta J_{ij}^{\phantom{0}}$ we have to deal with large quantities. For instance, for the nn interaction, we obtain (within GGA): $J_{1}^0 = -$$100.0$ meV and $\Delta J_{1}^{\phantom{0}} = 122.2$ meV, which strongly cancel each other. Obviously, once the Cl $3p$ states are explicitly included into the model, the strong coupling limit is not justified and MFT cannot be used: in GGA, MFT yields $J_{1}^0 = -$$7.6$ meV and $\Delta J_{1}^{\phantom{0}} = -$$1.1$ meV, which differ drastically from the parameters obtained in the exact approach. Furthermore, there is no more cancellation between $J_{1}^0$ and $\Delta J_{1}^{\phantom{0}}$ within MFT.

\par Then, we stick to the exact approach and search for the condition, when such cancellation would reproduce the experimental behavior of NaMnCl$_3$, namely the LA ground states with small $T_{\rm N}$ and small positive $\theta_{\rm CW}$. First of all, we have realized that, although the parameters ${\cal I}_{\mu}$ on the ${\rm T}$-sites are well defined, the ones on the ${\rm L}$-sites appear to be not. Generally, ${\cal I}_{\mu}$ is related to the exchange field $b_{\mu}^{z}$ and magnetization $m_{\mu}^{z}$ as ${\cal I}_{\mu} = -\frac{b_{\mu}^{z}}{m_{\mu}^{z}}$. However, there are several possibilities for evaluating these quantities in the TB model. For instance, $b_{\mu}^{z}$ can be associated with the site-diagonal elements of the TB Hamiltonian in the Wannier basis. Such $\hat{b}_{\mu}^{z} = \hat{H}_{\mu \mu}^{\uparrow} - \hat{H}_{\mu \mu}^{\downarrow}$ is the matrix in the subspace of the orbital indices. Then, in the spirit of DFT~\cite{KS}, corresponding site-diagonal magnetization matrix $\hat{m}_{\mu}^{z}$ should be obtained from the occupied eigenstates of $\hat{H}$. Hence, ${\cal I}_{\mu}$ can be defined as ${\cal I}_{\mu} = - \frac{{\rm Tr}_{L}(\hat{b}_{\mu}^{z})}{{\rm Tr}_{L}(\hat{m}_{\mu}^{z})}$. Another possibility (used above) is to employ the sum rule, $b_{\mu}^{z} = \sum_{\nu}[{\mathbb R}^{\uparrow \downarrow}(0)]_{\mu \nu}^{-1} m_{\nu}^{z}$~\cite{PRB2021}, in order to evaluate $b_{\mu}^{z}$ for the given set of magnetic moments and then use again the equation ${\cal I}_{\mu} = -\frac{b_{\mu}^{z}}{m_{\mu}^{z}}$. Nevertheless, we would like to emphasize that both definitions rely on assumptions. Particularly, the sum rule implies that the intersite matrix elements of the TB Hamiltonian $\hat{H}$ do not depend on spin, which is an approximation. On the other hand, the change $b_{\rm T}^{z}$ in the TB Hamiltonian, will affect not only $m_{\rm T}^{z}$ but also $m_{\rm L}^{z}$, even for fixed $b_{\rm L}^{z}$, thus making the definition ${\cal I}_{\rm L} = -\frac{b_{\rm L}^{z}}{m_{\rm L}^{z}}$ ambiguous.

\par The results are summarized in Table~\ref{tab:Stoner}.
\noindent
\begin{table}[b]
\caption{Effective Stoner parameters (in eV), as obtained in GGA for the antiferromagnetic (ferromagnetic) state by using site-diagonal elements of the tight-binding Hamiltonian and the sum rule.}
\label{tab:Stoner}
\begin{ruledtabular}
\begin{tabular}{ccc}
  method                                                                                      & ${\cal I}_{\rm Mn}$ & ${\cal I}_{\rm Cl}$ \\
\hline
${\cal I}_{\mu} = - \frac{{\rm Tr}_{L}(\hat{b}_{\mu}^{z})}{{\rm Tr}_{L}(\hat{m}_{\mu}^{z})}$  &   $0.88$ ($0.88$)   &   $1.40$ ($1.51$)   \\
sum rule                                                                                      &   $0.91$ ($0.90$)   &  $10.95$ ($8.75$)
\end{tabular}
\end{ruledtabular}
\end{table}
\noindent ${\cal I}_{\rm Mn}$ only weakly depends on the method and the magnetic state. The obtained values ${\cal I}_{\rm Mn} = 0.88 - 0.91$ eV are reasonable and consistent with ${\cal I}_{\rm Mn}=0.82$ eV reported for metallic fcc Mn~\cite{Janak}. On the other hand, depending on the definition and the magnetic state, ${\cal I}_{\rm Cl}$ varies from $1.4$ eV till $11$ eV (and even larger if one uses GGA$+$$U$ instead of GGA). The upper boundary for ${\cal I}_{\rm Cl}$, derived from the sume rule, may look unrealistically large. However, it should be understood that this ${\cal I}_{\rm Cl}$ is proportional to $\left[ {\mathbb R}^{\uparrow \downarrow} (\boldsymbol{q}) \right]^{-1}$, which nearly diverges in the case of NaMnCl$_3$. Furthermore, the concept of the Stoner splitting does not make sense for the totally filled Cl $3p$ shell in the ionic limit (contrary to the half-filled Mn $3d$ shell).

\par Thus, one can expect that the bare interaction $J_{ij}^0$, which does not explicitly depend on the Stoner coupling, is well defined. On the other hand, $\Delta J_{ij}^{\phantom{0}}$ is sensitive to the definition of ${\cal I}_{\rm Cl}$. In the following, we treat ${\cal I}_{\rm Cl}$ as an adjustable parameter, by varying it between the values obtained using the sum rule and site-diagonal elements of $\hat{H}$. Furthermore, we simulate the effect of the on-site $U$ by artificially changing the splitting between the majority ($\uparrow$) and minority ($\downarrow$) Mn $3d$ spin-states in the TB Hamiltonian. For the half-filled shell, this is equivalent to the change of the parameter ${\cal I}_{\rm Mn}$ and a constant shift of all Mn $3d$ states, which is controlled by the double-counting term in the GGA+$U$ approach. Since the form of this double counting is largely empirical, we consider the phenomenological procedure by fixing the positions of the occupied $\uparrow$-spin states and shifting upwards the unoccupied $\downarrow$-spin states, which accounts for the experimental situation in MnO~\cite{PRB1998}.

\par
\begin{figure}[b]
\begin{center}
\includegraphics[width=0.48\textwidth]{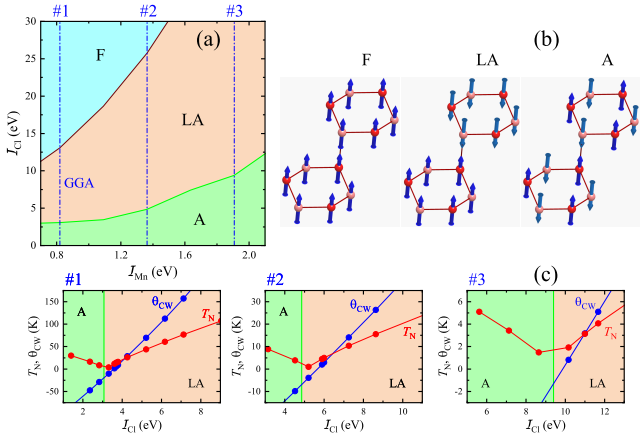}
\end{center}
\caption{(a) Phase diagram in the plane ${\cal I}_{\rm Mn}$-${\cal I}_{\rm Cl}$. (b) Ferromagnetic (F), layered antiferromagnetic (LA), and antiferromagnetic (A) structures. (c) The N\'eel temperature ($T_{\rm N}$) and Curie-Weiss temperature ($\theta_{\rm CW}$) as the function of ${\cal I}_{\rm Cl}$ for three characteristic values of ${\cal I}_{\rm Mn}$, which are indicated by the dot-dashed lines in (a). \#1 corresponds to GGA data.}
\label{fig.PD}
\end{figure}

 The basic idea is explained Fig.~\ref{fig.J}, in GGA. As expected, ${\cal I}_{\rm Cl}$ tends to strengthen the ferromagnetic interactions. For small ${\cal I}_{\rm Cl} = 1.4$ eV, all interactions are antiferromagnetic. Then, for the intermediate ${\cal I}_{\rm Cl} = 8.1$ eV, the nn interaction in the $xy$ plane, $J_1$, becomes ferromagnetic, while the interactions $J_2$, $J_4$, $J_5$, and $J_7$ between the planes remain antiferromagnetic. Finally, for large ${\cal I}_{\rm Cl} = 14.8$ eV, the interactions in and between the planes tends to be ferromagnetic. Thus, one can obtain the following phase diagram in the plane ${\cal I}_{\rm Mn}$-${\cal I}_{\rm Cl}$ (Fig.~\ref{fig.PD}).
\noindent

\par For each ${\cal I}_{\rm Mn}$ ($U$), the system undergoes the transition from the A to LA state and, then, to the F state with the increase of ${\cal I}_{\rm Cl}$. The parameters ${\cal I}_{\rm Cl}$ required to stabilize the F state are probably too large. Nevertheless, the A-LA transition takes place at more realistic values of ${\cal I}_{\rm Cl}$ ($\sim$$3.1$ in GGA, which further increases with the increase of $U$). Then, using the obtained parameters $J_{ij}$, we evaluate $T_{\rm N}$ (in RPA~\cite{PRM2019}) and $\theta_{\rm CW} = \frac{1+1/S}{3k_{\rm B}} \sum_{i} J_{i}$ (where $S=5/2$). The results are summarized in Fig.~\ref{fig.PD}c. We see that, in order to reproduce the experimental behavior of NaMnCl$_3$ (small $T_{\rm N}$ and small positive $\theta_{\rm CW}$), it is necessary to stay in the LA region, but close to the A-LA border. For instance,  for ${\cal I}_{\rm Mn} =$ $0.8$ (the GGA value), $1.4$, and $1.9$ eV, such a situation is realized around ${\cal I}_{\rm Cl} \sim$ $3.6$, $6.0$, and $11.0$ eV, respectively, yielding $(T_{\rm N},\theta_{\rm CW})=$ $(11.8,1.5)$, $(5.0,3.0)$, and $(3.1,3.2)$ K. Assuming $J_{\rm H}=0.8$ eV, ${\cal I}_{\rm Mn} =$ $0.8$, $1.4$, and $1.9$ eV, would correspond to the Coulomb $U=$ $0.9$, $3.6$, and $6.3$ eV, respectively. In all three cases, the required values of ${\cal I}_{\rm Cl}$ fall in-between two estimates based on the site-diagonal elements of $\hat{H}$ and the sum rule. Thus, irrespectively on the value of $U$, the experimental behavior of NaMnCl$_3$ can be formally reproduced by tuning ${\cal I}_{\rm Cl}$.

\par
\begin{figure}[t]
\begin{center}
\includegraphics[width=0.48\textwidth]{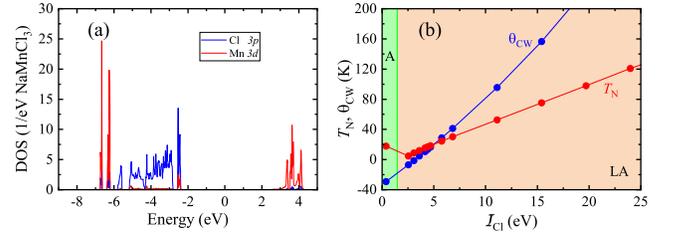}
\end{center}
\caption{Results of GGA$+$$U$ calculations: (b) Partial densities of Mn $3d$ and Cl $3p$ states. The zero energy is in the middle of the gap between occupied and empty states. (b) The N\'eel temperature ($T_{\rm N}$) and Curie-Weiss temperature ($\theta_{\rm CW}$) as the function of ${\cal I}_{\rm Cl}$.}
\label{fig.GGA+U}
\end{figure}

 Then, we turn to the brute force GGA$+$$U$ calculations with the parameters $U= 7.0$ eV and $J_{\rm H}= 0.8$ eV, which are typically extracted from constrained DFT calculations~\cite{AZA}. The main difference from the previous analysis is the double counting term, which is now taken in the standard form to reproduce the ionization potential and electron affinity of the Mn$^{2+}$ ions in the atomic limit~\cite{PRB1996}. This leads to the additional downward shift of the Mn $3d$ states, so that the occupied $\uparrow$-spin Mn $3d$ states are now located below the Cl $3p$ band (see Fig.~\ref{fig.GGA+U}a). Nevertheless, as far as the exchange interactions are concerned, this shift plays only a secondary role as the magnetic properties are mainly controlled by ${\cal I}_{\rm Cl}$, which strongly depends on the definition and varies from $2.5$ eV, if one uses site-diagonal elements of $\hat{H}$, till $45.5$ eV, if one uses the sum rule. The A-LA transition occurs near the lower boundary and the experimental behavior is captured reasonably well by ${\cal I}_{\rm Cl} \approx 3.6$ eV, yielding $T_{\rm N}=12.1$ K and $\theta_{\rm CW}=4.6$ K.
\noindent

\par Finally, taking into account the uncertainty with the choice of the parameters ${\cal I}_{\rm Cl}$ in the model, it is right to ask which is the realistic estimate of ${\cal I}_{\rm Cl}$ and which method is more suitable for capturing the experimental behavior of NaMnCl$_3$? For these purposes we evaluate effective exchange interactions by mapping the total energies obtained for several magnetic configurations onto the Heisenberg model (\ref{eq:Heis}). We consider five such configurations, which allow us extract three in-plane interactions $J_1$, $J_3$, and $J_6$, and one out-of-plane interaction $J_2$. Then, GGA (GGA$+$$U$) yields the following parameters (in meV): $J_1 = -1.1$ ($0.6$), $J_2 = -1.3$ ($-0.2$), $J_3 = -0.3$ ($0.1$), and $J_6 = -0.6$ ($0.2$). Thus, the ground state of NaMnCl$_3$ is expected to be of the A (LA) type with $T_{\rm N}=16.6$ ($10.8$) K and $\theta_{\rm CW}=-45.1$ ($13.3$) K in GGA (GGA$+$$U$). Definitely, GGA$+$$U$ does a better job, while plain GGA fails to reproduce the magnetic ground state of NaMnCl$_3$, though only by a tiny margin as the relatively small values of $T_{\rm N}$ and $\theta_{\rm CW}$ indicate that the situation is pretty close to the A-LA boundary. If we tried to reproduce such $T_{\rm N}$ and $\theta_{\rm CW}$ in the linear response theory, using Eq.~(\ref{eq:jexactM}) and the downfolding procedure for the ligand spins, we would have to use ${\cal I}_{\rm Cl} \approx 2.6$ ($4.4$) eV in GGA (GGA$+$$U$). These values are close to the lower boundaries of ${\cal I}_{\rm Cl}$, evaluated via site-diagonal elements of $\hat{H}$, which still need to be increased by about 80\% in order to reproduce the results of the total energy calculations.

\par \textit{Summary and Conclusions}. -- The response theory for interatomic exchange interactions is the powerful modern tool as it provides not only quantitative estimates, but also a deep microscopic understanding of the origin of these interactions. Nevertheless, in order to describe these interactions properly in NaMnCl$_3$ and other $90^{\circ}$-exchange systems, it is essential to take into account the polarization of the ligand states -- the contribution, which was typically ignored in most of the approximate techniques based on MFT. We explain the origin of this contribution, which naturally incorporates the dependence of $J_{ij}$ on not only the Coulomb $U$ on the Mn sites, but also the Stoner coupling ${\cal I}_{\rm Cl}$ on the ligand sites, as was long advocated by the phenomenological GKA rules. Yet, the main obstacle for truly \textit{ab initio} description along this line is the definition of the parameter ${\cal I}_{\rm Cl}$ for the nearly filled Cl $3p$ shell, which does not seem to be unique.

\par As the next important development of the linear response technique, it would be interesting to include the relativistic spin-orbit interaction, which is responsible for anisotropic Dzyaloshinskii-Moriya interactions (DMI) and exchange anisotropy. Particularly, our study clearly demonstrate the crucial role of the ligand $p$ states in the exchange processes. However, there are many examples when the same ligand states play a key role in developing the large anisotropy and DMI, as demonstrated in CrI$_3$~\cite{CrI3_Lado,CrI3_Liu} and Cr$_2$Ge$_2$Te$_6$~\cite{PCCP}. Furthermore, the ligand assistant exchange anisotropy in honeycomb layered cobaltites may lead to the realization of Kitaev spin liquid state, similar to Na$_2$IrO$_3$ and $\alpha$-RuCl$_3$~\cite{Liu_PRL}. In this respect, NaCoCl$_3$ seems to be particularly promising candidate if it could be synthesized.

\par S.V.S would like to thank Robert Cava for stimulating discussions. The work was supported by the project RSF 20-62-46047 of the Russian Science Foundation in the part of conventional DFT calculations and the program 122021000038-7 (Quantum) of the Russian Ministry of Science and High Education in the part of construction and analysis of microscopic models of interatomic exchange interactions.

\end{document}